\documentclass[twoside]{hs04}
\usepackage{pstricks}
\def\runauthor{Giacomo Sguazzoni}
\def\shorttitle{Higgs and Beyond Standard Model searches at LEP}
\usepackage{graphicx}
\usepackage{pstricks}
\usepackage[figuresright]{rotating}
\def \pb        {{\rm \, pb}}
\def \fb        {{\rm \, fb}}
\def \ipb       {{\rm \, pb^{-1}}}

\def \GeV       {{\rm \, GeV}}

\def \GeVcc     {\GeV/c^2}

\def\ga{\mathrel{\raise.3ex\hbox{$>$\kern-.75em\lower1ex\hbox{$\sim$}}}}
\def\la{\mathrel{\raise.3ex\hbox{$<$\kern-.75em\lower1ex\hbox{$\sim$}}}}

\newcommand {\gtrsim}
     {\,\raisebox{-0.6ex}{$\stackrel{\textstyle>}{\textstyle\sim}$}\,}

\newcommand {\bfell}      {\ell\kern-0.4em
                           \ell\kern-0.4em
                           \ell\kern-0.4em
                           \ell }
\newcommand {\obfell}     {\overline{\ell}\kern-0.4em
                           \overline{\ell}\kern-0.4em
                           \overline{\ell}\kern-0.4em
                           \overline{\ell}}
\newcommand {\bfH}      {\; {\cal H}\kern-0.5em \kern-0.4em
                           {\cal H}\kern-0.5em \kern-0.4em
                           {\cal H}\kern0.1em }
\newcommand {\obfH}     {\; \overline{\cal H}\kern-0.5em \kern-0.4em 
                           \overline{\cal H}\kern-0.5em \kern-0.4em 
                           \overline{\cal H}\kern0.1em }
\def \b             {{\mathrm b}}
\def \t             {{\mathrm t}}
\def \charm         {{\mathrm c}}
\def \d             {{\mathrm d}}
\def \u             {{\mathrm u}}
\def \e             {{\mathrm e}}
\def \q             {{\mathrm q}}

\def \h             {{\mathrm h}}
 
\def \f             {{\mathrm f}} 
\def \A             {{\mathrm A}}

\def \W             {{\mathrm W}}
\def \H             {{\mathrm H}}
\def \Z             {{\mathrm Z}}

\def \P             {{\mathrm P}}
\def \G             {{\mathrm G}}
\newcommand {\ho}         {{\h^0}}
\newcommand {\Ho}         {{\H^0}}
\newcommand {\Ao}         {{\A^0}}
\newcommand {\Hpm}        {{\H^\pm}}
\newcommand {\clsb}       {{\mathrm CL_{\rm s+b}}}
\newcommand {\clb}        {{\mathrm CL_{\rm b}}}
\newcommand {\dM}         {\Delta M}

\newcommand {\sfe}     {{\tilde{\f}}}
\newcommand {\sfL}     {{\tilde{\f}_{\mathrm L}}}
\newcommand {\sfR}     {{\tilde{\f}_{\mathrm R}}}

\newcommand {\sneu}    {{\tilde{\nu}}}

\newcommand {\seR}     {{\mathrm{\tilde{e}_{R}}}}

\newcommand {\st}      {{\mathrm{\tilde{\tau}}}}
\newcommand {\stR}     {{\mathrm{\tilde{\tau}_{R}}}}

\newcommand {\smR}     {{\mathrm{\tilde{\mu}_{R}}}}

\newcommand {\sto}     {{\tilde{\mathrm{t}}}}

\newcommand {\sbot}    {{\tilde{\mathrm{b}}}}

\newcommand {\snu}     {{\tilde{\nu}}}

\newcommand {\neu}     {{\chi}}

\newcommand {\thstop} {\mathrm{\theta_{\tilde{t}}}}
\newcommand {\thsbot} {\mathrm{\theta_{\tilde{b}}}}

\newcommand {\tanb}{\tan\beta}
\newcommand {\qq}    {{\q \overline{\q}}}
\newcommand {\bb}    {{\b \overline{\b}}}

\newcommand {\ee}    {{\e ^+ \e ^-}}

\newcommand {\pmiss}   {{P\!\!\!\,\!/ }}
\newcommand {\emiss}   {{E\!\!\!\,\!/ }}
\begin{document}

\title{HIGGS AND BEYOND STANDARD MODEL\\
SEARCHES AT LEP}

\author{GIACOMO~SGUAZZONI}

\address{University of Pisa and INFN,\\
via Buonarroti 2, I-50126 Pisa, Italy\\
E-mail: giacomo.sguazzoni@cern.ch}

\maketitle

\abstracts{
Extensive searches for Higgs bosons and other new phenomena predicted
by extensions of the Standard Model have been performed at LEP. A
summary is given reviewing the principal aspects and presenting a
selection of results.
}

\section{Introduction}

The LEP analyses devoted to searches benefitted from the
impressive performance of the accelerator that, during a 
decade, provided $\ee$ collisions for an integrated
luminosity of $900\ipb$ per experiment at centre-of-mass energies
ranging from $88$-$95\GeV$ (LEP 1) and $130$-$209\GeV$ (LEP
2). Four detectors were operating: ALEPH, DELPHI, L3 and OPAL, all
designed for precision physics at the Z peak and beyond,
featuring large covering angle, good particle identification 
for leptons, photons and b quarks, and good jet and energy flow
reconstruction. New phenomena could in fact be searched
indirectly, via precision  measurements, or directly, trying to
identify unexpected $\ee$ events. 

Precision measurements show that SM works very well and constraints 
can be derived to new contributions to the observables. As a general
example, the agreement of the measured Z width with the SM prediction
excludes new particles with a non negligible coupling to the Z if
their masses are smaller than $\sim M_\Z/2$.

Direct searches rely on the highest centre-of-mass energy and the
large integrated luminosity of the LEP 2 phase.  
The typical LEP-combined
excluded cross section is of the order of $\sim0.01-0.1\pb$ for
pair-produced new particles up to masses around half the
center-of-mass energy. 


Three main streams can be recognized: searches for the Higgs bosons,
searches for Supersymmetry (SUSY), 
and 
searches for a large variety of other non-SUSY extensions of the
SM. The results, based on the full high-energy data sample, are
presented in the form of 95\% C.L. exclusion domains in the space of
the relevant parameters, since no excess has been observed. When
available, the LEP SUSY Working Groups combinations, based on the
outcomes from all experiments (ADLO), are
reported~\cite{LEPHIGGSWG,LEPSUSYWG,LEPEXWG}.  

\section{Higgs bosons searches}


In the Standard Model (SM), gauge bosons and fermions obtain
their masses interacting with the vacuum Higgs fields. Associated with
this description is the existence of 
massive scalar particles, the Higgs bosons, yet to be
discovered. 

The minimal SM requires one Higgs field doublet and predicts a single
neutral Higgs boson, H. Indirect experimental bounds for the SM Higgs
boson mass ($m_\H = 96^{+60}_{-38}\GeVcc$, $m_\H <219\GeVcc$ at the 95\%
C.L.) are obtained from fits to precision measurements
of electroweak observables~\cite{LEPEWWG}. 
The SM Higgs boson can be produced mainly via the
  Higgs--strahlung process $\ee\to\Z\H$, up to Higgs boson masses
  below the ``kinematic wall'', approximately given by
  $\sqrt{s}+m_\Z$. The contribution from the $\W\W\H$ fusion channel,
  $\ee\to\H\nu_{\e}\overline{\nu}_{\e}$ and $\ee\to\H\ee$, normally
  negligible, plays also some role close to the kinematic limit. The
  main decay channels are $\bb$ and $\tau^+\tau^-$ pairs, the
  latter marginally contributing ($\sim 8\%$).

Given the production mechanism and decays channels, four topologies,
corresponding to different final states, 
have been searched for: the four-jet topology,
$\ee\to(\H\to\bb)(\Z\to\qq)$, that occurs with a branching 
ratio of about 60\%; the missing
energy topology, $\ee\to(\H\to\bb)(\Z\to\nu\overline{\nu})$ that occurs with a
branching ratio of about 17\%; the leptonic topology,
$\ee\to(\H\to\bb)(\Z\to\ell^+\ell^-)$ ($\ell=\e$, $\mu$) whose small
branching ratio ($\sim6\%$) is balanced by the intrinsic low
background; the tau topology, $\ee\to(\H\to\tau^+\tau^-)(\Z\to\qq)$ or
$\ee\to(\H\to\bb)(\Z\to\tau^+\tau^-)$, that occur with a branching ratio
of about 10\% in total. The Higgs boson is reconstructed by tagging the
b jets from its decay whereas the remaining particle system is required to
be kinematically compatible with a $\Z$ decay. All these criteria are
often optimally exploited by using neural-network techniques.

By using the likelihood ratio method, a LEP-wide test-statistic
$Q({\rm data}|m_\H)$ is built from the outcomes of these searches
by all experiments~\cite{LEPHIGGSWG}. For any given $m_\H$, 
it allows the compatibility of the selected events with the
signal-plus-background (``s+b'') hypothesis or with the
background-only (``b'') hypothesis to be evaluated. $Q({\rm data}|m_\H)$
is compared with the expected pdf's of $Q({\rm 
  s+b},m_\H)$ and $Q({\mathrm b},m_\H)$ (both computed by using MC
  simulations) and two probabilities $\clsb$ and $1-\clb$, known 
as {\em confidence levels}, are derived. They represent the
  probabilities that the outcome of a new experiment is more ``s+b''-like
  or ``b''-like, respectively, than the outcome represented by the
  set of selected events.
\begin{figure}[!t]
  \begin{center}
    \begin{picture}(0,0)
      \put(0,0){\mbox{(a)}}
    \end{picture}
    \includegraphics[width=0.38\textwidth]{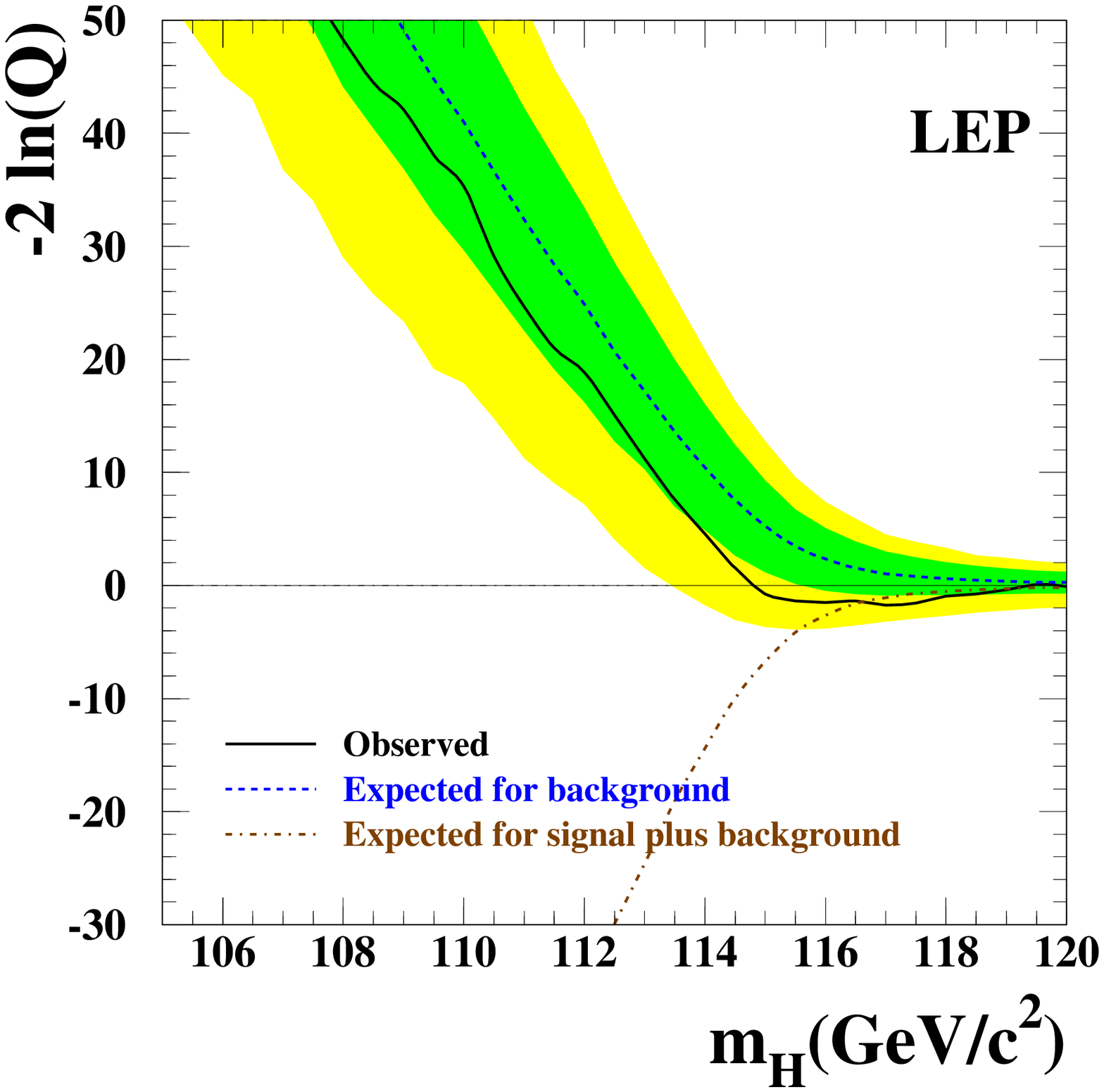} 
    \hskip 0.04\textwidth
    \begin{picture}(0,0)
      \put(0,0){\mbox{(b)}}
    \end{picture}
    \includegraphics[width=0.38\textwidth]{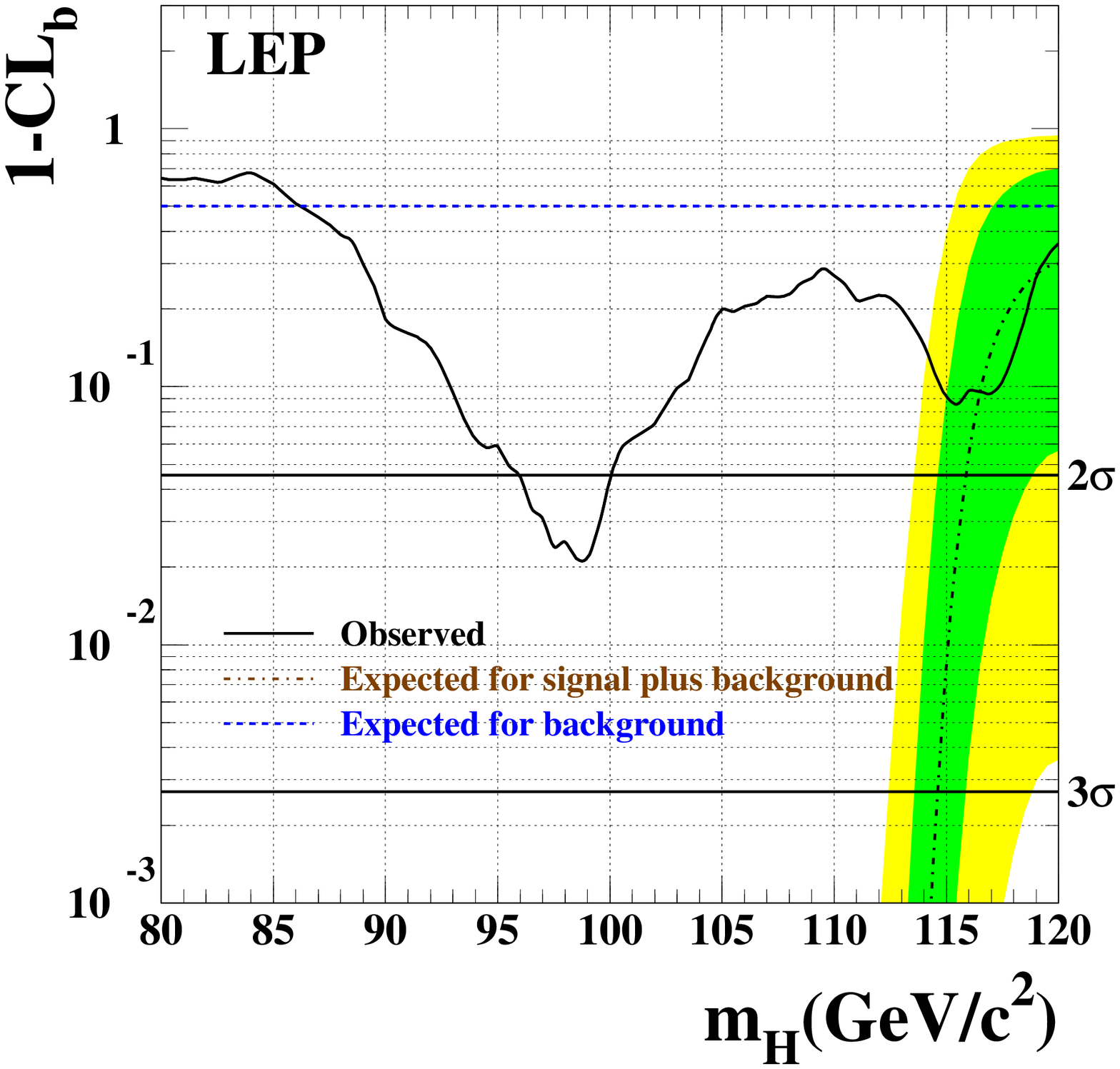} 
    \caption[*]{(a) The test-statistic as a function of  
      $m_\H$. The dashed line represents the expectation in the ``b''
      only hypothesis with the $1\sigma$ and $2\sigma$ probability
      bands; the solid line is the observed results, whereas the
      dotted line is the median result expected in presence of a
      $m_\H=115\GeVcc$ signal.
      (b) $1-\clb$ as a function of $m_\H$; the line at 0.5 is median
      result in the absence of a signal, the solid curve is observed
      result and the dashed curve is median result expected for a
      signal.
      }
      \label{fig:higgs}
  \end{center}
\end{figure}

The LEP combined test-statistic $Q$ as a function of the hypothetical $m_\H$
is shown in Fig.~\ref{fig:higgs}(a), in the more 
convenient and normally used form of $-2\ln Q$ that can be simply
written as a sum of observed event weights. The negative broad minimum
crossing the ``s+b'' expected curve at
$m_\H\sim115\GeVcc$ favours the signal hypothesis for an Higgs boson mass
around that value. The significance is limited to
$2.3\sigma$ overall, but, as the expected curves indicate, it is
compatible with the sensitivity achievable for that mass
range. The ``b'' confidence level, visible in
Fig.~\ref{fig:higgs}(b), is 8\%, with a corresponding ``s+b'' confidence
level of $37\%$. This excess concentrates into the ALEPH data set at
the level of $2.8\sigma$ and it is mainly due to four-jet candidates
with clean b tags and kinematic properties~\cite{alephhiggs1}. It is not
confirmed neither excluded from the other experiments but has been
demonstrated to be robust and stable~\cite{alephhiggs2}.
However, since no unambiguous signal has been observed, a lower limit
on the SM Higgs boson mass of $114.4\GeVcc$ has been set, slightly
below the expected value ($115.3\GeVcc$).

Except simplicity there is no a priori reason justifying the presence
of only one Higgs doublet in the SM. The two-Higgs-doublet models (2HDM)
represent the simplest extensions of the SM Higgs
sector. LEP searches have also addressed their
phenomenology~\cite{LEPHIGGSWG}.

In the context of a general 2HDM, the Higgs sector comprises five
physical Higgs bosons: two neutral CP-even scalars, $\ho$ and $\Ho$
(with $m_\ho<m_\Ho$), one CP-odd scalar, $\Ao$, and two charged
scalars, $\Hpm$. Their masses are free parameters and the
choice of the couplings between the Higgs bosons and the fermions
determines the type of the 2HDM model: in the Type-I models the
quarks and leptons only couple to the second Higgs doublet; in the
Type-II models the first Higgs doublet couples only to down-type
fermions and the second Higgs doublet couples only to up-type fermions.

The $\ho$ and $\Ao$ bosons are expected to be predominantly produced 
via the Higgs-strahlung, $\ee\to\Z\ho$ and $\Z\Ho$, and
the associated production, $\ee\to\ho\Ao$ and 
$\Ho\Ao$. The decay properties of the Higgs bosons,
while quantitatively different depending on models and their
parameters, maintain a certain similarity with the SM case:
$\bb$ and $\tau^+\tau^-$ are the relevant decays channels but
only if the cascade decays (i.e. $\ho\to\Ao\Ao$) are not kinematically
allowed and dominant. These similarities very often allow the SM Higgs
boson searches to be used for 2HDM Higgs bosons. Nevertheless special
situations may occur for which specific analyses need to be developed:
searches for $\ho\Z$ and $\ho\Ao$ independently on the flavour of the
decay channel~\cite{LEPHIGGSWG}, and searches for $\ho\Z$
independently on the decay channel~\cite{opalhiggsdecayind} are just
two examples.

\begin{figure}[!t]
  \begin{center}
    \begin{picture}(0,0)
      \put(0,0){\mbox{(a)}}
    \end{picture}
    \includegraphics[width=0.32\textwidth]{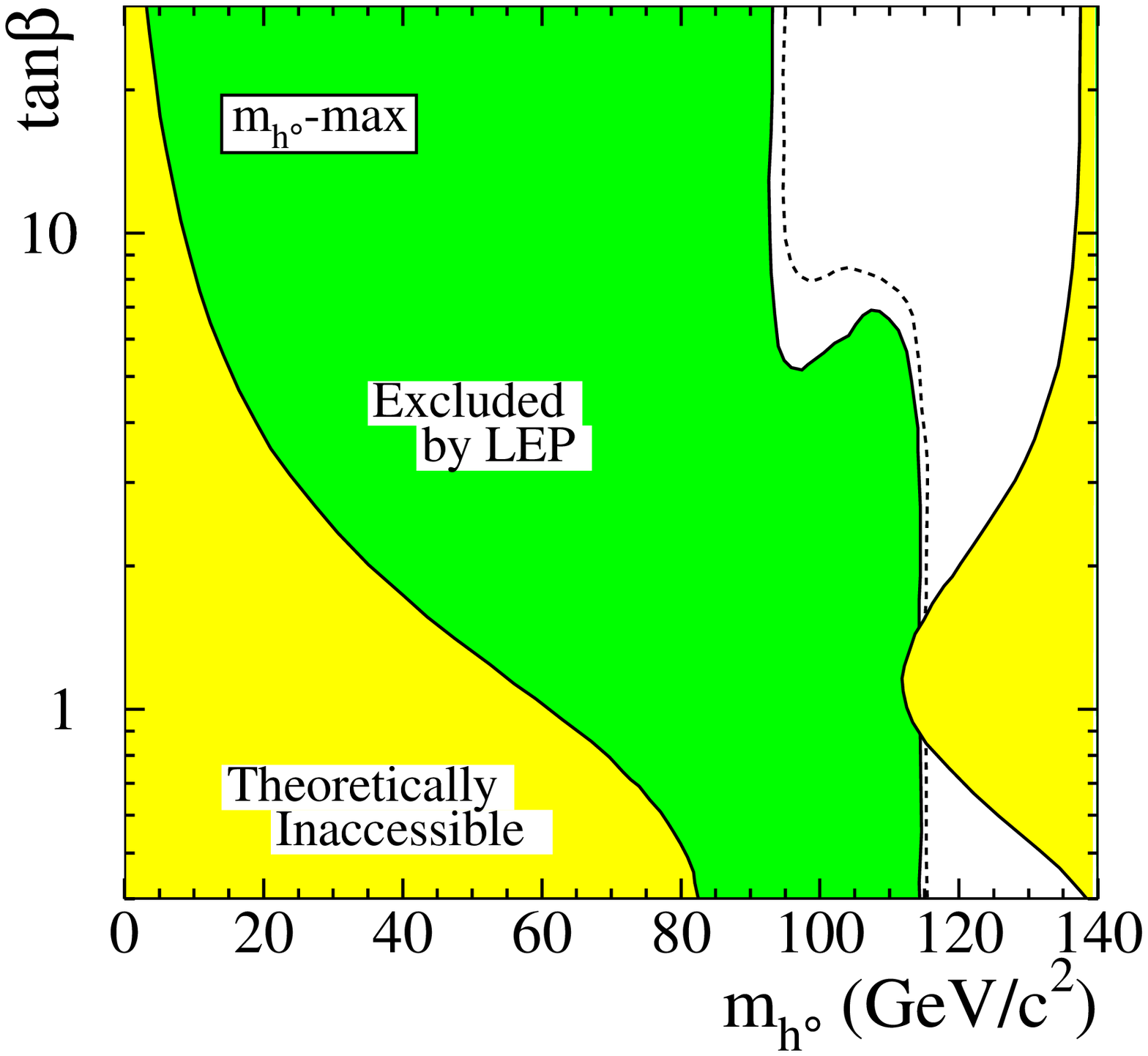} 
%
    \begin{picture}(0,0)
      \put(0,0){\mbox{(b)}}
    \end{picture}
    \includegraphics[width=0.32\textwidth]{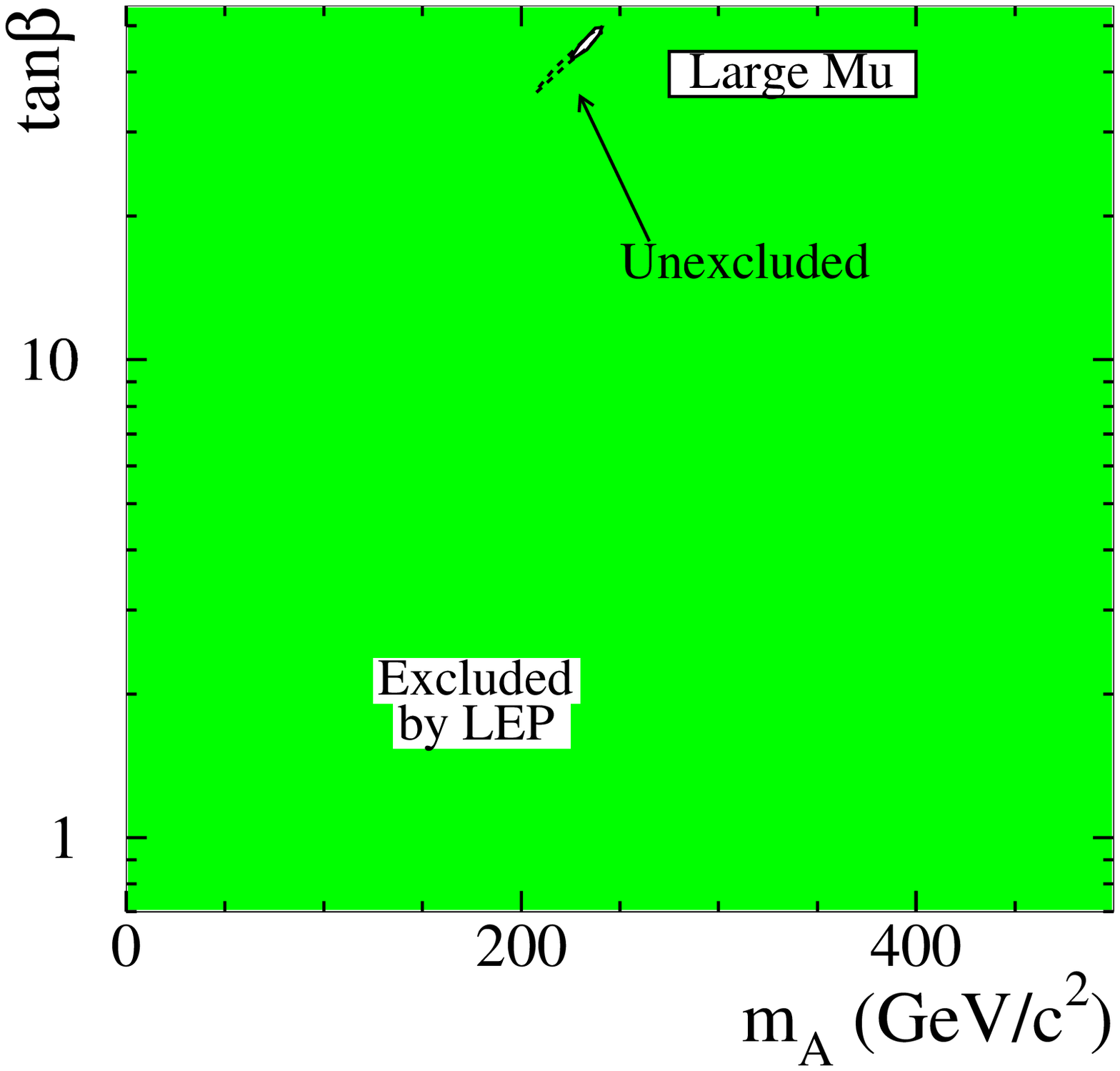} 
%
    \begin{picture}(0,0)
      \put(0,0){\mbox{(c)}}
    \end{picture}
    \includegraphics[width=0.295\textwidth]{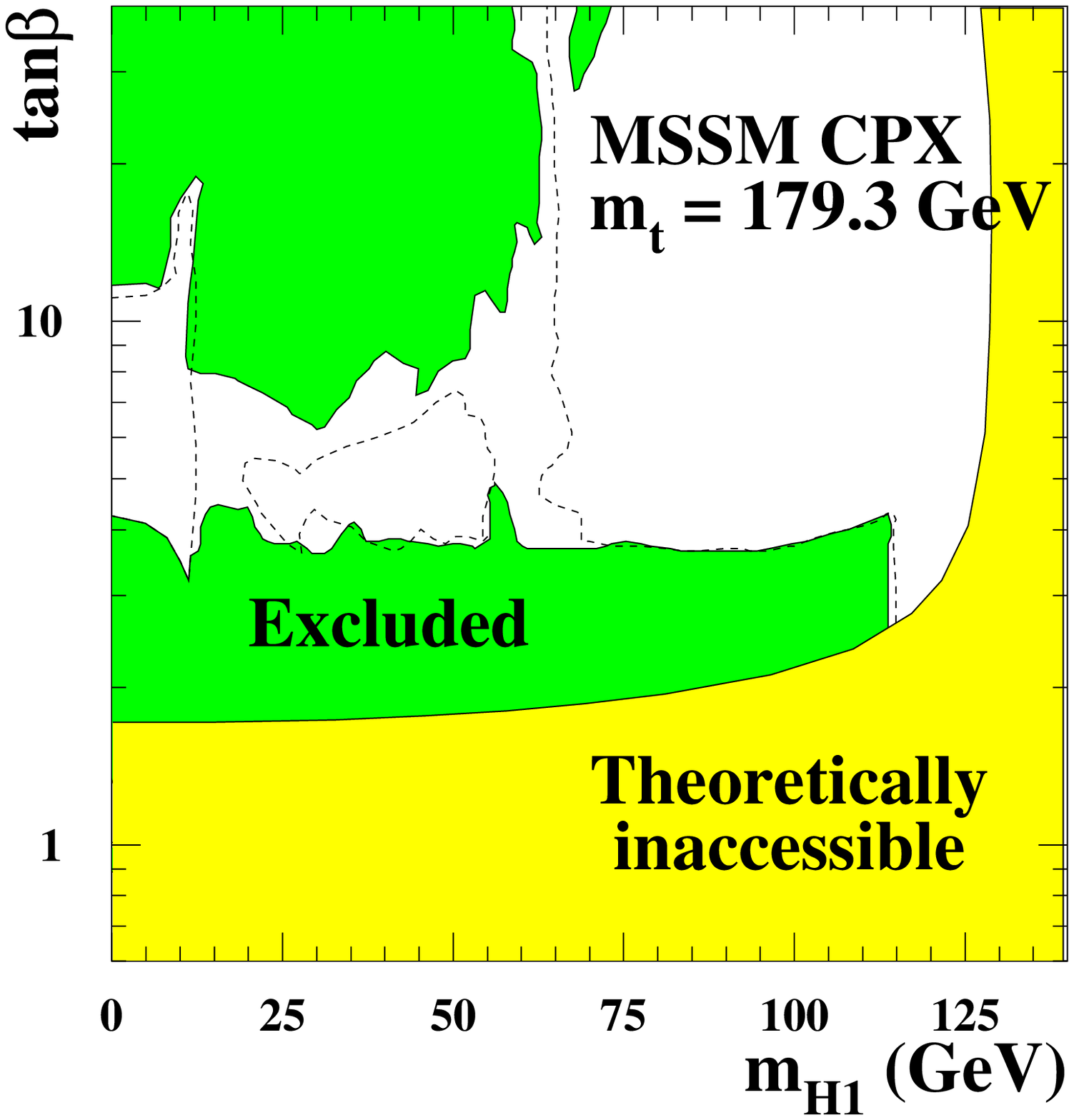} 
    \caption[*]{The MSSM exclusions for various benchmarks: (a) the
      $m_\ho$-max scenario in the ($m_\ho$, $\tanb$) projection, with
      $m_\t = 179.3 \GeVcc$.; (b) the large-$\mu$ scenario in
      the ($m_\Ao$, $\tanb$) projection; (c) the CP-violating CPX
      scenario, for $m_\t = 179.3 \GeVcc$, in
      the ($m_{\H_1}$, $\tanb$) projection. The dashed lines indicate
      the boundaries of the expected exclusions.} 
      \label{fig:mssmhiggs}
  \end{center}
\end{figure}

The most important implementation of a 2HDM model, Type II, is the
Minimal Supersymmetric Model or MSSM (see Sec.~\ref{sec:susy}). Due to
its importance, the MSSM is used to challenge the LEP capability to
detect 2HDM Higgs bosons by interpreting the search results within special MSSM
benchmark scenarios. The ``$m_\ho$-max'' scenario, designed to
maximise the theoretical $m_\ho$ upper bound, provides a wider
parameter space and therefore more conservative exclusion limits, as shown
in Fig.~\ref{fig:mssmhiggs}(a). In the ``large-$\mu$'' scenario the lightest
Higgs boson is everywhere kinematically accessible but its detection is
expected to be difficult since the $\bb$ decay mode, on which 
most of the searches rely, is strongly suppressed; nevertheless,
thanks to the flavour independent searches, this scenario is almost
entirely excluded, as visible in Fig.~\ref{fig:mssmhiggs}(b).

All these scenarios are CP-conserving (CPC). In general, however,
the neutral Higgs bosons $\ho$, $\Ao$ and $\Ho$ can mix into three
states $\H_1$, $\H_2$, $\H_3$ with not defined CP quantum numbers
giving rise to CP-violating models (CPV) that are also studied. Their
phenomenology remain similar 
either for production ($\ee\to\H_i\Z$, $\ee\to\H_i\H_j$, $i$,
$j=1,2,3$, $i\neq j$) and decay ($\bb$, $\tau^+\tau^-$ and
cascades $\H_j\to\H_i\H_i$, $j>i$). A benchmark model, know as ``CPX'',
has been chosen to maximise the phenomenological differences with
respect to the CP-conserving scenarios. A CPX exclusion example is
shown in Fig.~\ref{fig:mssmhiggs}(c).

General models can also yield to other interesting Higgs boson
signals that not necessarily are contemplated into the considered MSSM
scenarios: few examples are given here~\cite{LEPHIGGSWG}.
The ``invisible'' Higgs boson decays into an undetectable final
state, as a neutralino pair. The selections designed to address its
strahlung production rely on the visible recoil system from the 
$\Z$ decay and allow mass limits of the order of $\gtrsim110\GeVcc$ to
be set, assuming a SM like production cross section.
Within the MSSM, charged Higgs bosons $\H^\pm$ are normally heavier than
other Higgs bosons; nevertheless specific selections exist by which mass
limits of the order of $\sim77-78\GeVcc$ are set for all possible
decay channels. 
A Higgs boson that does not couple to fermions, known as
``fermiophobic'', is possible in 2HDM models of Type I. Specific
analyses for such Higgs boson, consisting of a search for strahlung
production and decay into photons, allow mass limits around $110\GeVcc$
to be set for a decay branching ratio into photons exceeding $\sim 
10\%$.

\section{SUSY searches}
\label{sec:susy}

Theories with Supersymmetry (SUSY) are the most promising extensions
of the Standard Model (SM). The simplest version is
the Minimal Supersymmetric Model (MSSM), which contains the minimal
number of additional particles. The scalar fermions or {\em
  sfermions}, $\sfL$ and 
$\sfR$, are the partners of the left- and right-handed SM fermions and
mix to form the mass eigenstates. The mixing angle $\theta_{\sfe}$ is
so defined that $\sfe\!=\!\sfL\cos\theta_{\sfe}\!+\!\sfR\sin\theta_{\sfe}$ is
the lightest sfermion. In general mixing is relevant for the third
family, while $\sfe\!\equiv\!\sfR$ otherwise.
The SM gauge boson as well as MSSM Higgs bosons states have fermionic
super-partners, {\em gauginos} and {\em higgsinos}. The neutral
higgsinos and gauginos mix into four mass
eigenstates, the {\em neutralinos} $\chi$, $\chi_2$, $\chi_3$,
$\chi_4$
($M_{\chi_4}\!>\!M_{\chi_3}\!>\!M_{\chi_2}\!>\!M_{\chi}$). The charged
gauginos and higgsinos mix into two mass eigenstates, the {\em
  charginos} $\chi^{\pm}$ and $\chi^{\pm}_{2}$
($M_{\chi^\pm_2}\!>\!M_{\chi^\pm}$).

The lepton and baryon number conservation is normally embedded into
SUSY models through the ``R-parity'' conservation. The LSP (Lightest
Supersymmetric Particle) is stable and must be also neutral and weakly
interacting to fit the cosmological observations. Within the standard
MSSM the LSP is the lightest neutralino $\chi$ or, less likely, the
sneutrino, $\sneu$. At LEP the sparticles are pair produced and the
decay brings to final states containing at least one LSP. 

Except few pathological cases, sparticle pair production leads to the typical
acoplanar particles topology due to missing energy ($\emiss$) and
momentum ($\pmiss$) from escaping LSP's. The energy of the visible
system is related to the mass difference between the sparticle $\widetilde{\P}$
and the LSP ($\dM\!=\!M_{\widetilde{\P}}\!-\!M_\mathrm{LSP}$).
The acoplanar topologies studied cover each type of visible final
state (leptons, hadronic jets, $\gamma$'s).

\begin{figure}[!t]
  \begin{center}
    \begin{picture}(0,0)
      \put(0,0){\mbox{(a)}}
    \end{picture}
    \includegraphics[width=0.36\textwidth]{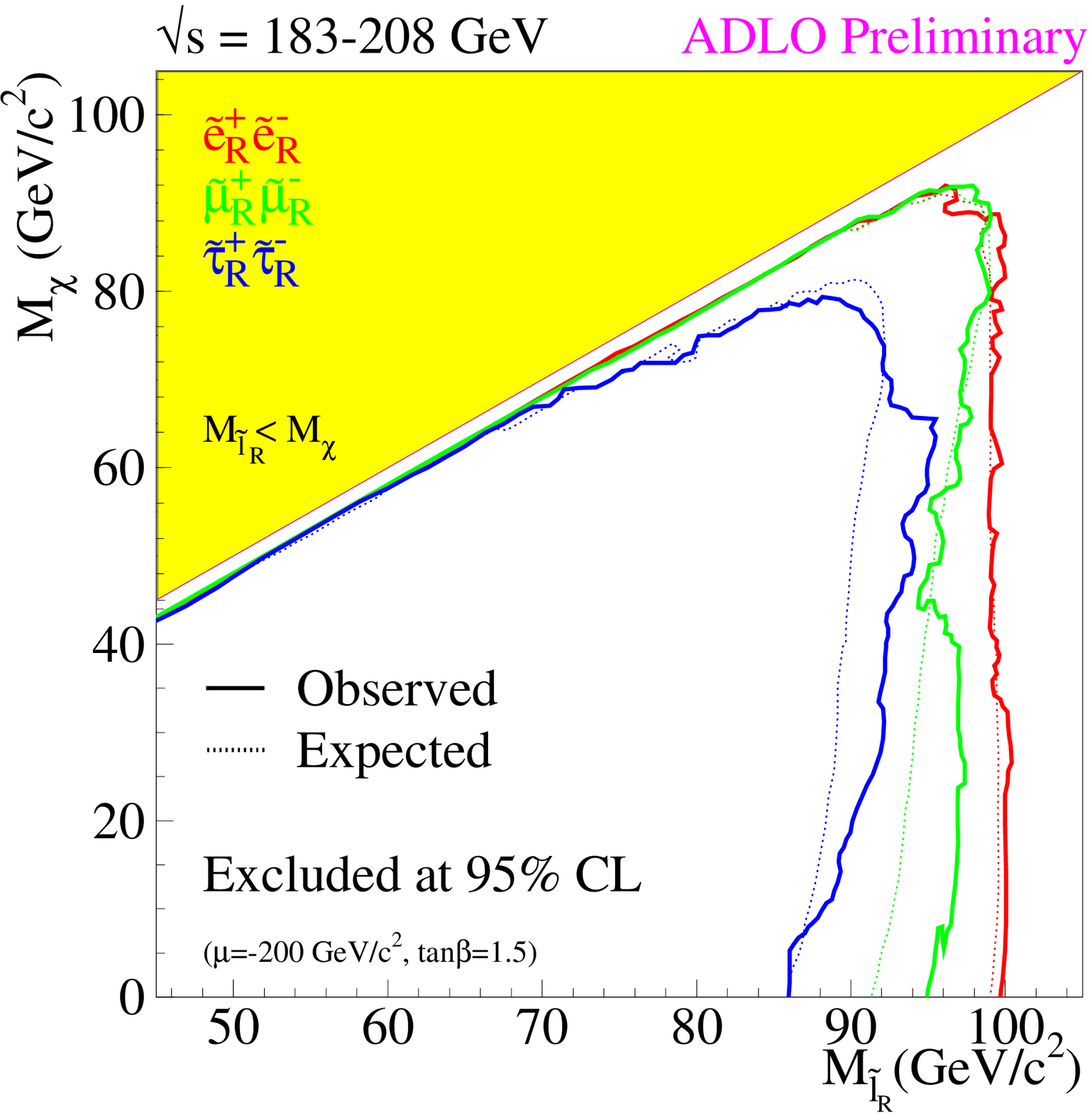} 
    \hskip 0.04\textwidth
    \begin{picture}(0,0)
      \put(0,0){\mbox{(b)}}
    \end{picture}
    \includegraphics[width=0.31\textwidth]{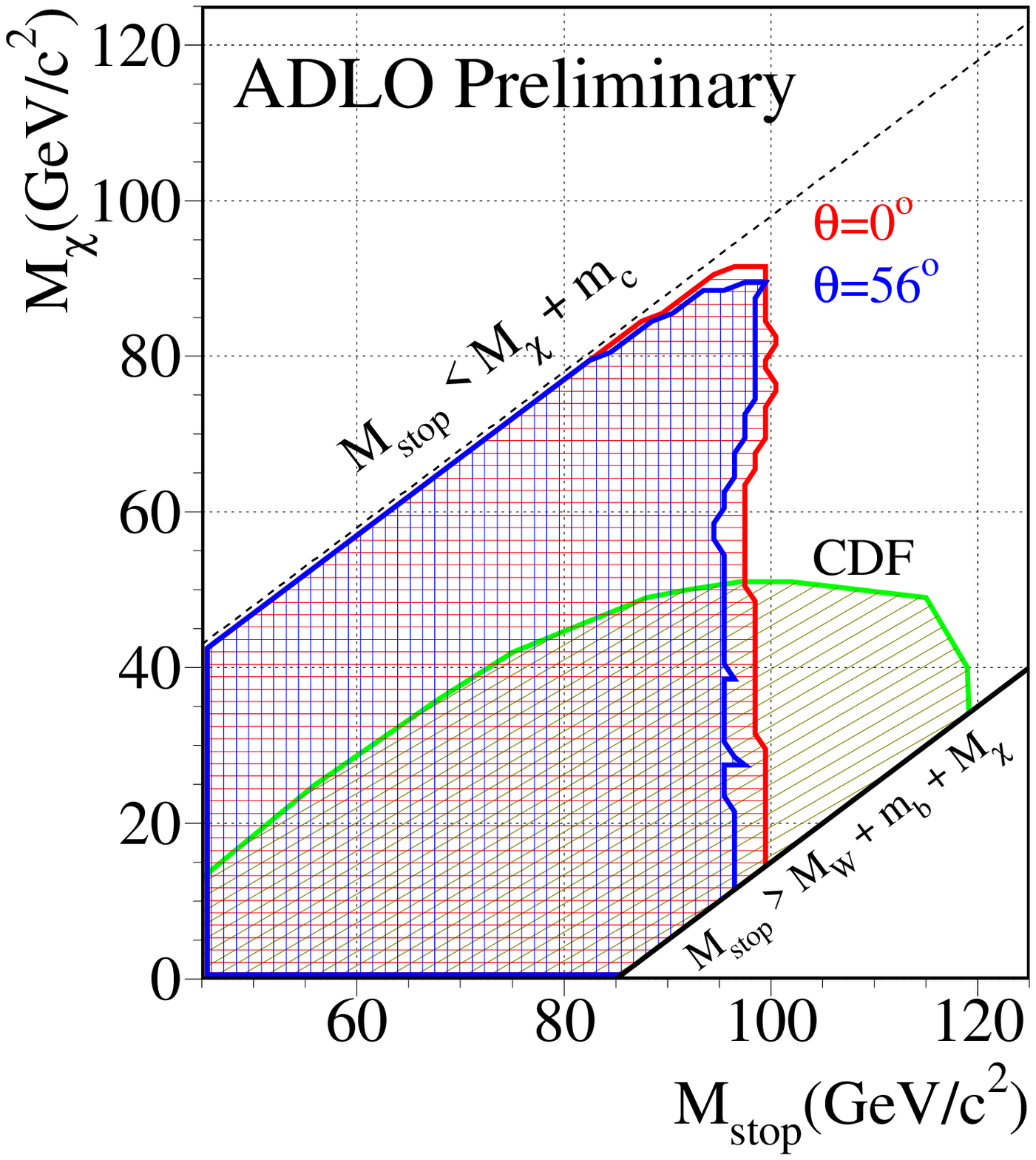} 
    \caption[*]{LEP SUSY Working Group results for sfermions: (a)
    slepton mass exclusion plot; (b) stop mass exclusion plot in case
    of $\sto\!\to\!\charm\neu$ decay for minimal 
    ($\thstop\!=\!56^\circ$) and maximal production cross section
    ($\thstop\!=\!0^\circ$).} 
    \label{fig:sfermions}
  \end{center}
\end{figure}

The analyses for slepton signals
($\ee\!\to\!\tilde{\ell}^{+}\tilde{\ell}^{-}$, $\tilde{\ell}\!\to\!\ell\neu$) 
search for acoplanar leptons by using
the powerful lepton and tau identification of LEP
detectors, and the LEP combined cross section upper limits range from $10$ 
to $60\fb$. The resulting mass lower limits are $100\GeVcc$,
$94\GeVcc$ and $86\GeVcc$ for $\seR$, $\smR$ and $\stR$
respectively, valid for $\dM\!>\!10\GeVcc$, as shown in
Fig.~\ref{fig:sfermions}(a)~\cite{LEPSUSYWG}.

The production of a squark pair results into an acoplanar
jet topology. These hadronic events can be selected by using event
variables and requiring $\emiss$ and $\pmiss$. 
In case of $\ee\!\to\!\sto\bar{\sto}$, $\sto\!\to\!\charm\neu$, the
mass lower limit is $94\GeVcc$ for $\dM\!>\!10\GeVcc$ and any mixing,
as visible in Fig.~\ref{fig:sfermions}(b)~\cite{LEPSUSYWG}. Further
specialized selections are used for other squark processes: b-tagging
is effective for $\ee\!\to\!\sbot\bar{\sbot}$, $\sbot\!\to\!\b\neu$,
allowing a limit of $92\GeVcc$ to be set ($\dM\!>\!10\GeVcc$, any
$\thsbot$); leptons are required in case of
$\ee\!\to\!\sto\bar{\sto}$, $\sto\!\to\!\b\ell\sneu$, leading to a
mass lower limit of $95\GeVcc$ ($\dM\!>\!10\GeVcc$, any $\thstop$).
The stop decay $\sto\!\to\!\b\neu\f_{\u}\bar{\f}_{\d}$ leads to a
multi-body final state topology addressed by a dedicated ALEPH
selection~\cite{ALEPH_squark_209}. As an 
example, assuming the decay $\sto\!\to\!\b\chi\W^*$, the result is
$M_\sto\!>\!77\GeVcc$ ($\dM\!>\!10\GeVcc$, any $\thstop$).
ALEPH analyses also consider the case in which a stop quasi-degenerate
with the LSP acquires a sizeable lifetime and
hadronizes~\cite{ALEPH_stopldm}. This scenario has been excluded
searching for long-lived heavy hadrons and an absolute stop mass lower
limit of $63 \GeVcc$ has been set for any $\thstop$, any branching ratio and
any $\dM$~\cite{ALEPH_squark_209}.

Topologies with two or more visible fermions in the final state plus
$\emiss$ and $\pmiss$ are expected in case of charginos and neutralinos
production~\cite{lsplimits}.  
The processes are of the type 
$\ee\!\to\!\chi_{i>1}\chi$ and $\ee\!\to\!\chi_{i>1}\chi_{j>1}$ with
$\chi_{i>1}\!\to\!\chi\f\bar{\f}$, and $\ee\!\to\!\chi^{+}\chi^{-}$
with $\chi^\pm\!\to\!\chi\f_{\rm u}\f'_{\rm d}$. 
Cross section upper limits of $\sim\!0.1$--$0.3\pb$ are obtained by the
LEP-wide outcome of dedicated selections~\cite{LEPSUSYWG}.

Topologies with photon(s) can be very powerful in detecting new
phenomena~\cite{LEPSUSYWG},
in general being sensitive to pair produced sparticles radiatively
decaying into the LSP. Within the MSSM this case applies to neutralino
production processes like $\ee\!\to\!\chi_2\chi_2$ and
$\ee\!\to\!\chi_2\chi$ with 
$\chi_2\!\to\!\chi\gamma$. In this hypotheses the cross section upper
limits range between $10\fb$ and $0.1\pb$ depending on the
process~\cite{LEPSUSYWG}. 

The negative results of the search for sparticle production
can be translated into constraints on the parameter space
in the context of specific SUSY models. This method allows the
exclusions to be extended to sparticles otherwise not accessible,
either because invisible, as the LSP, either because too heavy to be
produced~\cite{lsplimits}.

A widely accepted framework is the constrained MSSM (CMSSM). 
The unification of masses and 
couplings at the GUT scale allow the EW scale phenomenology to be set by
few parameters: $\tanb$, the ratio of the vacuum 
expectation values of the two Higgs doublets; $\mu$, the Higgs sector
mass parameter; $M_2$, the EW scale common gaugino    
mass; $m_0$, the GUT scale common scalar mass; the
trilinear couplings $A_\f$, that enter in the
prediction of the sfermion mixing and are generally set to fit the
no-mixing hypothesis.

\begin{figure}[!t]
  \begin{center}
    \begin{picture}(0,0)
      \put(0,0){\mbox{(a)}}
    \end{picture}
    \includegraphics*[width=0.31\textwidth,clip]{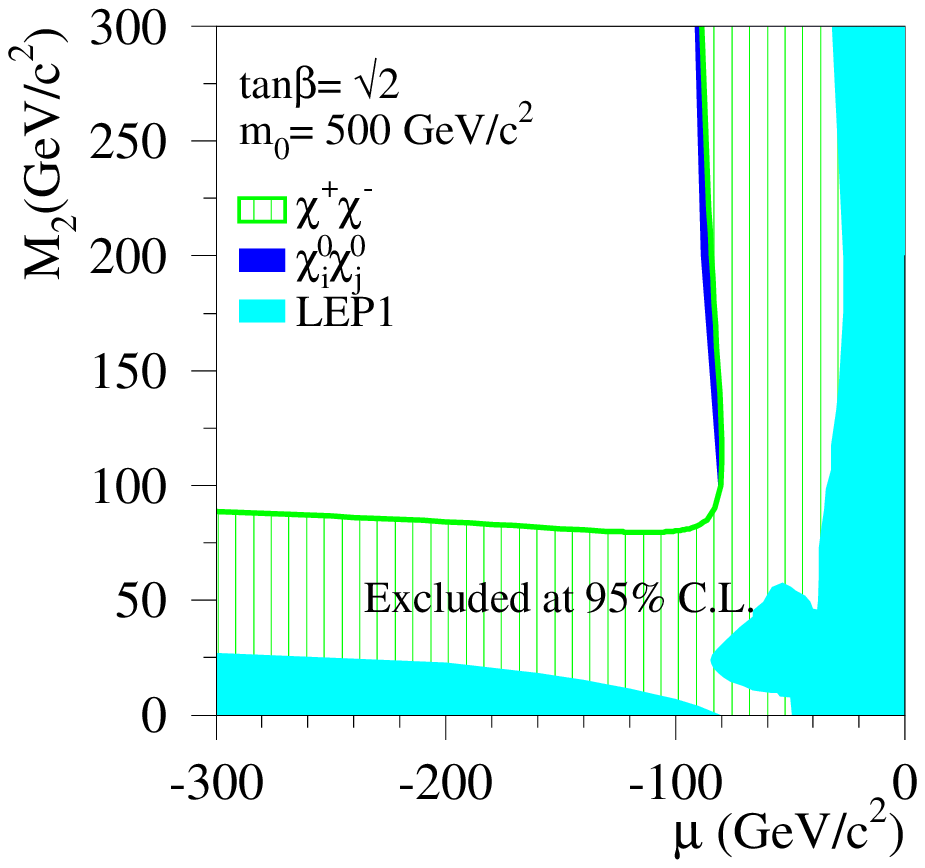} 
    \hskip 0.01\textwidth
    \begin{picture}(0,0)
      \put(0,0){\mbox{(b)}}
    \end{picture}
    \includegraphics*[width=0.30\textwidth]{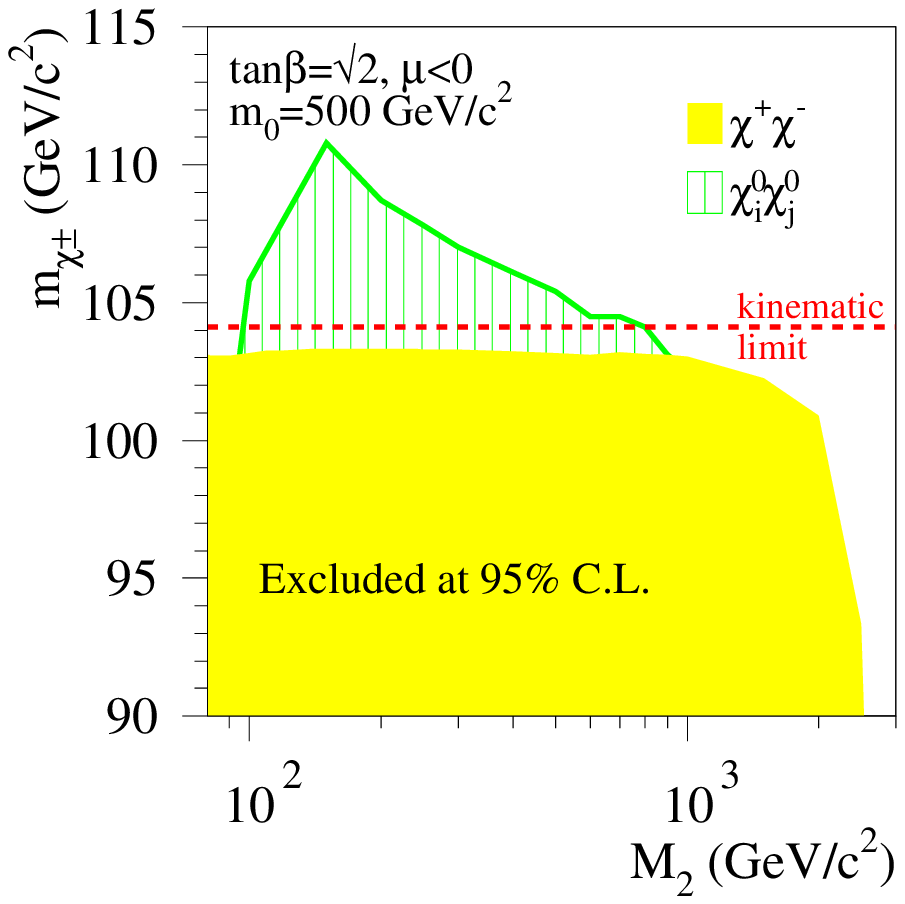} 
    \hskip 0.01\textwidth
    \begin{picture}(0,0)
      \put(0,0){\mbox{(c)}}
    \end{picture}
    \includegraphics[width=0.31\textwidth]{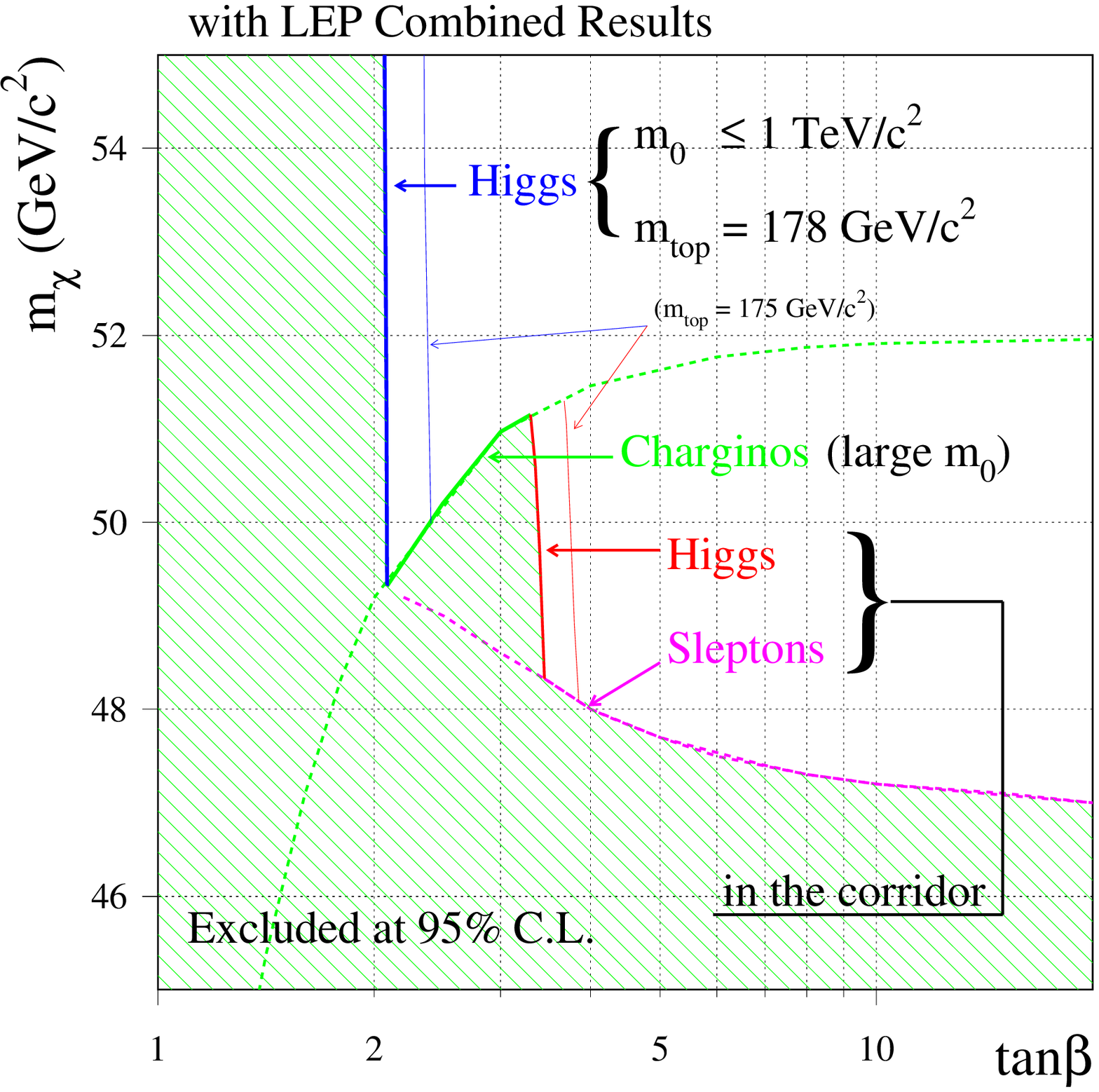} 
    \caption[*]{(a) ALEPH exclusion in the $M_2$ versus negative $\mu$
      plane for $\tanb\!=\!\sqrt{2}$ and $m_0\!=\!500\GeVcc$; (b)
      ALEPH chargino mass lower limits for $\mu\!<\!0$;
      (c) Absolute lower limit on the LSP mass in the CMSSM as a
      function of $\tanb$.
    } 
    \label{fig:gauginos}
  \end{center}
\end{figure}

The negative outcome of charginos and neutralinos searches can be
used to exclude regions in the $(\mu,M_2)$ plane, as shown, as an
example, in Fig.~\ref{fig:gauginos}(a) in which the sleptons are assumed
to decouple (i.e. large $m_0$). Figure~\ref{fig:gauginos}(b) shows how
neutralino searches allow chargino exclusions to be improved for small
$\tanb$ and $\mu\!<\!0$. If the sleptons are lighter (small $m_0$
values), the chargino and neutralino cross sections decrease for the
enhancement of negative-interfering slepton-exchange diagrams. The
consequent loss of sensitivity is recovered by slepton searches in
such a way that lower mass limits on gauginos and other sparticles as 
$\seR$ or $\snu$ could be set.
Among these, the most important is the LSP limit, i.e. the mass lower
limit on $\chi$. The LSP mass lower limits from LEP experiments fall
around $36-39\GeVcc$ and are set for $\tanb\!\sim\!1$~\cite{lsplimits}.

The LEP mass lower limits on the Higgs boson mass $m_\ho$ can be also
used to further exclude small $\tanb$ ranges. Roughly, this 
just derives from the MSSM tree-level relation $m_\ho\!<\!m_\Z|\!\cos
\beta|$. However, the details of the exclusion depend on $M_2$, $m_0$
and the stop mass because of the large radiative corrections to $m_\ho$.
Adding the Higgs constraints the LSP mass lower limit substantially improves (up
to $\sim\!47\GeVcc$) and moves towards high $\tanb$, as shown in
Fig.~\ref{fig:gauginos}(c)~\cite{LEPSUSYWG}.

The robustness of the LSP limit has been checked with respect to the
mixing effects in the third family, neglected in the above discussion. 
A stau getting light for mixing may be mass degenerate with the LSP,
making the chargino decays into staus difficult to detect.
Dedicated selections for
$\chi^\pm\!\to\!\st\nu_\tau\!\to\!\tau\chi\nu_\tau$ with soft taus,  
$\ee\!\to\!\chi_2\chi$ and $\ee\!\to\!\chi_2\chi_2$
with $\chi_2\!\to\!\tau\tau\chi$, and for chargino production in
association with an ISR photon ($\ee\!\to\!\chi^+\chi^-\gamma$) 
allow to solve this problem. The LSP limits reported above have
been demonstrated to hold by using this studies, extended also considering
the mixing configurations for $\st$, $\sto$ and $\sbot$ that can
be explored by setting $A_\tau$, $A_\t$ and $A_\b$ to zero.
\begin{figure}[!b]
  \begin{center}
    \begin{picture}(0,0)
      \put(0,0){\mbox{(a)}}
    \end{picture}
    \includegraphics*[width=0.33\textwidth,clip]{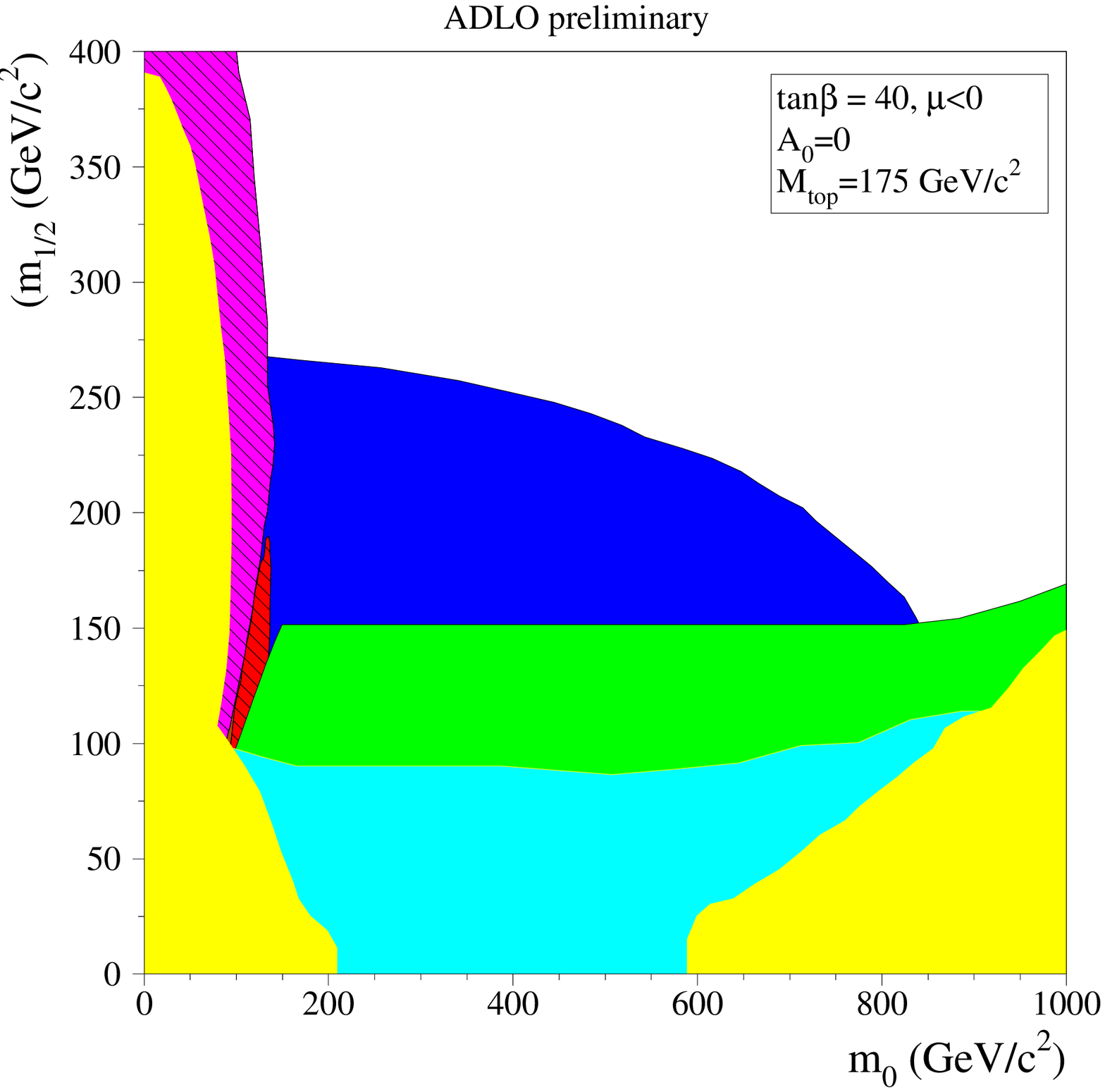} 
    \hskip 0.04\textwidth
    \begin{picture}(0,0)
      \put(0,0){\mbox{(b)}}
    \end{picture}
    \includegraphics*[width=0.33\textwidth]{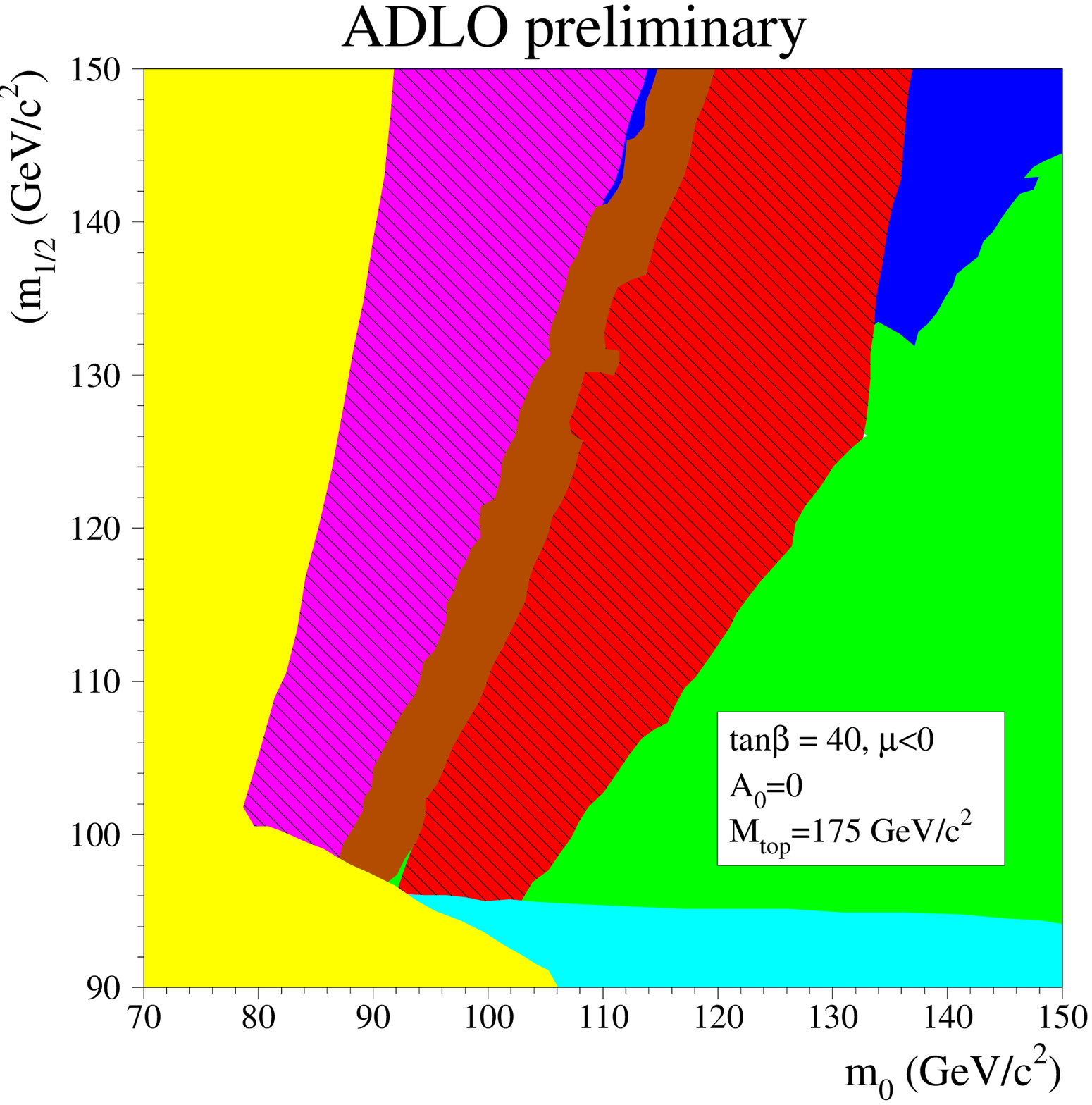} 
    \begin{picture}(0,0)
      \put(-244,22){\mbox{1}}
      \put(-170,22){\mbox{1}}
      \put(-100,70){\mbox{1}}
      \put(-210,25){\mbox{2}}
      \put(-50,16){\mbox{2}}
      \put(-205,40){\mbox{3}}
      \put(-30,55){\mbox{3}}
      \put(-62,50){\mbox{\white\bf\small 4}}
      \put(-215,55){\mbox{\white\bf 5}}
      \put(-25,96){\mbox{\white\bf 5}}
      \put(-243,102){\mbox{\white\bf\small 6}}
      \put(-75,90){\mbox{\white\bf\small 6}}
      \put(-73,55){\mbox{\white\bf\small 7}}
    \end{picture}
    \caption[*]{(a) LEP combined exclusion domains in the mSUGRA $m_{1/2}$
      versus $m_0$ plane for $\tanb\!=\!40$, $A_0\!=\!0$ and
      $\mu\!<\!0$; a peculiar area is zoomed in (b) to show the
      interplay between selections. Numbers indicate the search
      excluding the corresponding area: (1) theoretically not allowed,
      (2) LEP1, (3) chargino, (4) stau and selectron, (5) Higgs, (6)
      $\chi\to\st$ cascade and (7) heavy stable charged 
      particles.} 
    \label{fig:mSUGRA}
  \end{center}
\end{figure}

The results have been also interpreted within an even more constrained
version of the CMSSM, usually referred to as Minimal Supergravity (mSUGRA).
The relevant parameters are: $\tanb$, the sign of $\mu$ and $m_0$;
 $m_{1/2}$, the GUT scale common gaugino mass, that replaces
$M_2$; $A_0$, the GUT scale common trilinear coupling. 

On top
of LEP1 exclusions and theory-forbidden regions, small $m_0$ and $m_{1/2}$
areas are constrained from sleptons and gaugino searches, respectively. Higgs
boson searches are also effective, even in the large $\tanb$
range. As an example, Fig.~\ref{fig:mSUGRA}(a) illustrates $m_{1/2}$
versus $m_0$ excluded domains for $\tanb\!=\!40$, $\mu\!<\!0$ and
$A_0\!=\!0$. The zoomed area of Fig.~\ref{fig:mSUGRA}(b) focuses on
the pathological region in which, for 
the mixing, $\st$ and $\chi$ are almost degenerate and the
selections for stau-cascades and stable staus have to be used. 
The resulting mSUGRA LSP mass lower limits lie between $52$ and $59\GeVcc$,
depending on the top mass, and turn out to be $\sim\!8$--$9\GeVcc$
lower if $A_0$ is allowed to assume values other than
zero~\cite{LEPSUSYWG}.

Gauge-mediated SUSY breaking (GMSB) models are
characterized by the following distinctive features: the LSP is always
the gravitino $\tilde{\G}$, and the next-to-LSP (NLSP) is, in general,
either the lightest $\chi$ or a slepton; the NLSP decay length could be even 
comparable or larger than the detector dimension depending on
parameters. Within GMSB, double- or single-photon final state may occur
in case of neutralino pair production and radiative decay into
gravitinos ($\ee\to\chi\chi\to\tilde{\G}\tilde{\G}\gamma\gamma$). A
complete addressing of GMSB topologies requires, in general, the use
of searches for acoplanar particles as well as searches for kinks
and/or impact parameter for the intermediate lifetime range, and for
heavy stable charged particles for the long lifetime
range~\cite{LEPSUSYWG}.

Supersymmetry does not necessarily require R-parity conservation (RPC)
and R-parity violation (RPV) can be introduced still complying with
low energy limits on barion and lepton number
conservation. Two of the main characteristics of SUSY
processes at LEP are heavily affected: sparticles can also be singly
produced and mainly, since LSP is no more stable, the decay final
state is build up of standard model particles with many leptons and/or
quarks, and the missing mass signature is generally lost. The huge
amount of possible and complex final states foreseen in this scenario
are addressed by many LEP analyses with results comparable to the RPC
case. The study of these decays allowed to test very
peculiar topologies, otherwise unsearched for.

\section{Search for extra-dimensions: an example of ``exotic'' searches}
\label{exotic}

The searches of beyond SM phenomena that do not fit within
supersymmetric models are generally said as ``exotic''. The huge
amount of work done by LEP collaboration on this topic can not be
covered here~\cite{LEPEXWG}. As a representative example the search
for extra dimension is briefly reviewed.

In order to address the hierarchy problem, a new class of
theories assume the Standard Model to be confined into a
four-dimensional hypersurface (brane), whereas the gravitational
fields are also allowed to propagate in extra dimensions
inside the full space-time (bulk). Two different extra dimensions
scenarios have been studied at LEP: the ADD and the
Randall-Sundrum scenarios. 
\begin{figure}[!t]
  \begin{center}
    \begin{picture}(0,0)
      \put(0,0){\mbox{(a)}}
    \end{picture}
    \includegraphics[width=0.36\textwidth]{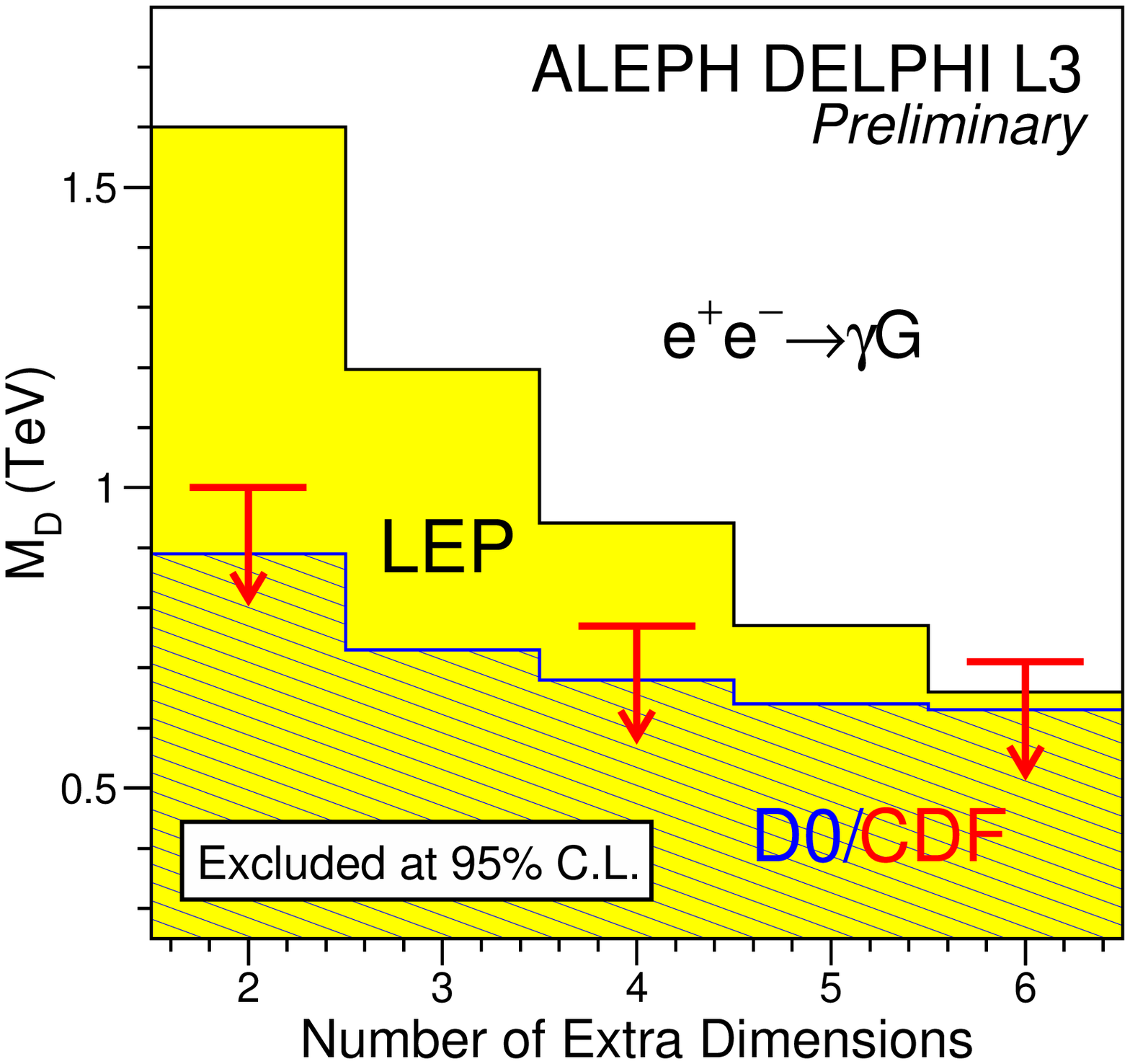} 
    \hskip 0.04\textwidth
    \begin{picture}(0,0)
      \put(0,0){\mbox{(b)}}
    \end{picture}
    \includegraphics[width=0.36\textwidth]{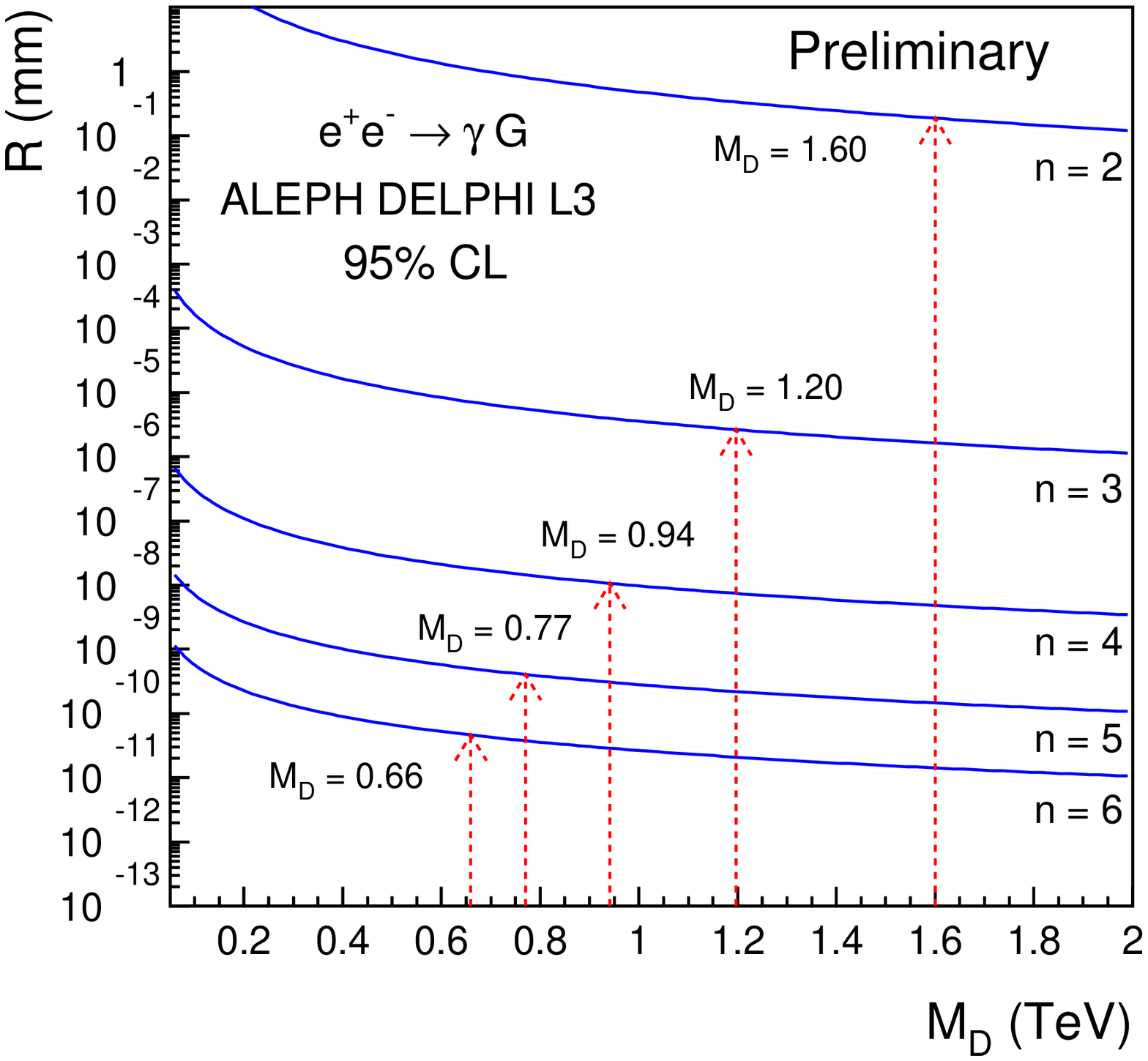} 
    \caption[*]{
      (a) The exclusion contours in the $M_D$ vs $n$ plane for the
      graviton-photon emission at LEP (ADL combined). Current limits
      from D0 and CDF are also shown.
      (b) The radii of the extra dimensions, $R$, as functions of the
      gravity scale $M_D$, for $n=2-6$. Arrows indicate the upper limits
      on $R$.
      }
      \label{fig:extradim}
  \end{center}
\end{figure}

In the ADD scenario, the $n>1$ extra dimensions are assumed to be
compactified, normally on a torus with radius $R$. The fundamental
gravitational scale $M_D$ is related to the Planck scale $M_{\rm
  Pl}\sim10^{19}\GeV$ through $R$ ($M_{\rm Pl}^2\sim M_D^{2+n} R^n$),
and can be lowered to the TeV range with the extra dimension size
being as large as a millimeter. Gravitational fields propagating in
the bulk can be expressed as a series of states known as a
Kaluza-Klein towers. An observer trapped on the brane sees the
graviton modes propagating in extra dimensions as massive spin-2
neutral particles which can couple to the SM fields on the
brane. In the presence of large extra dimensions, events with a single
photon and missing energy could be enhanced by $\ee\to\G\gamma$
processes where the graviton G escapes detection. Since 
no indication of a signal has been observed~\cite{LEPEXWG}, limits on
the scale of gravity $M_D$, shown in Fig.~\ref{fig:extradim}(a), are
derived. These can be converted into upper limits on R, as shown in
Fig.~\ref{fig:extradim}(b).

\begin{figure}[!t]
  \begin{center}
    \begin{picture}(0,0)
      \put(0,0){\mbox{(a)}}
    \end{picture}
    \includegraphics[width=0.36\textwidth]{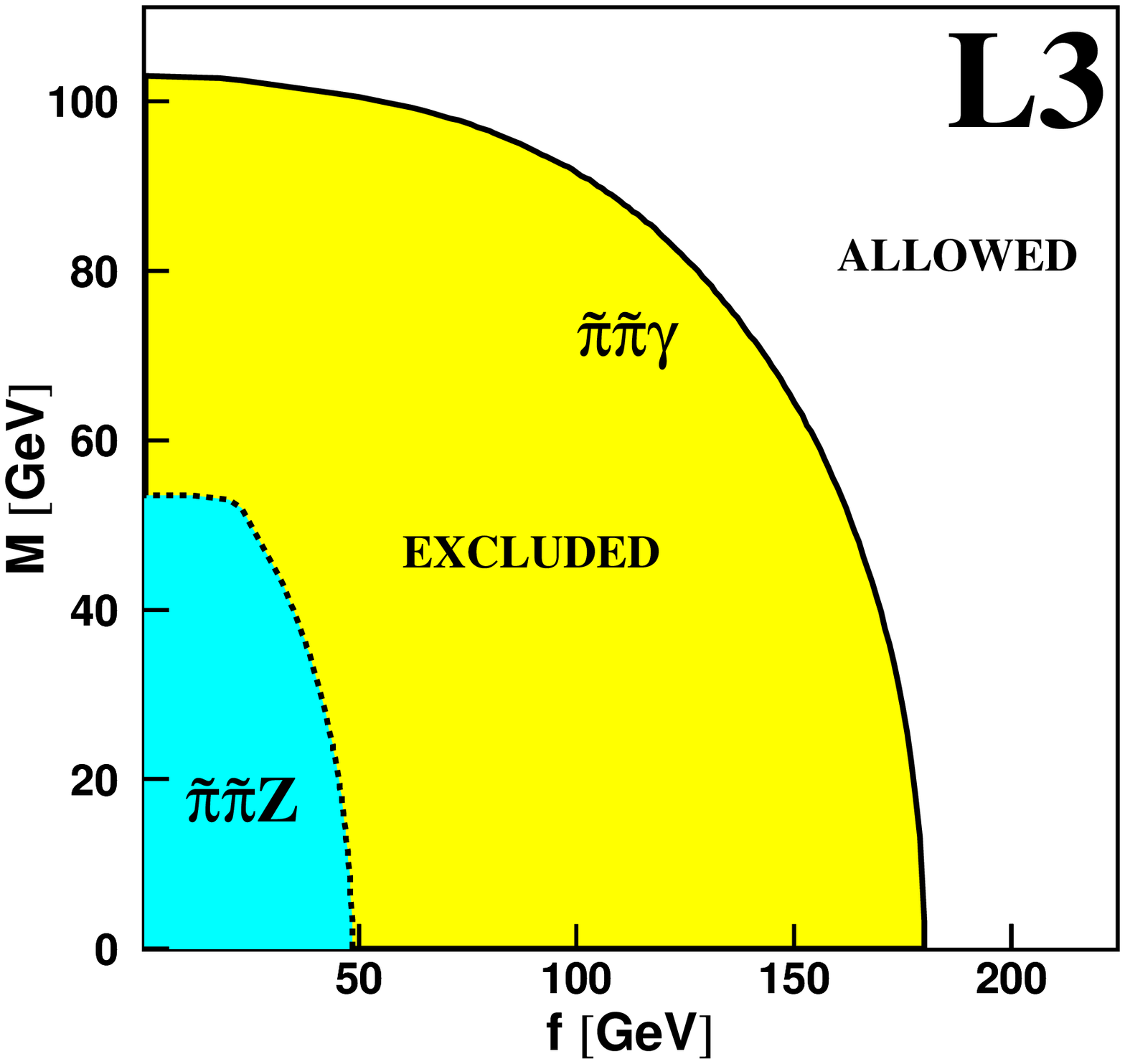} 
    \hskip 0.04\textwidth
    \includegraphics*[width=0.45\textwidth]{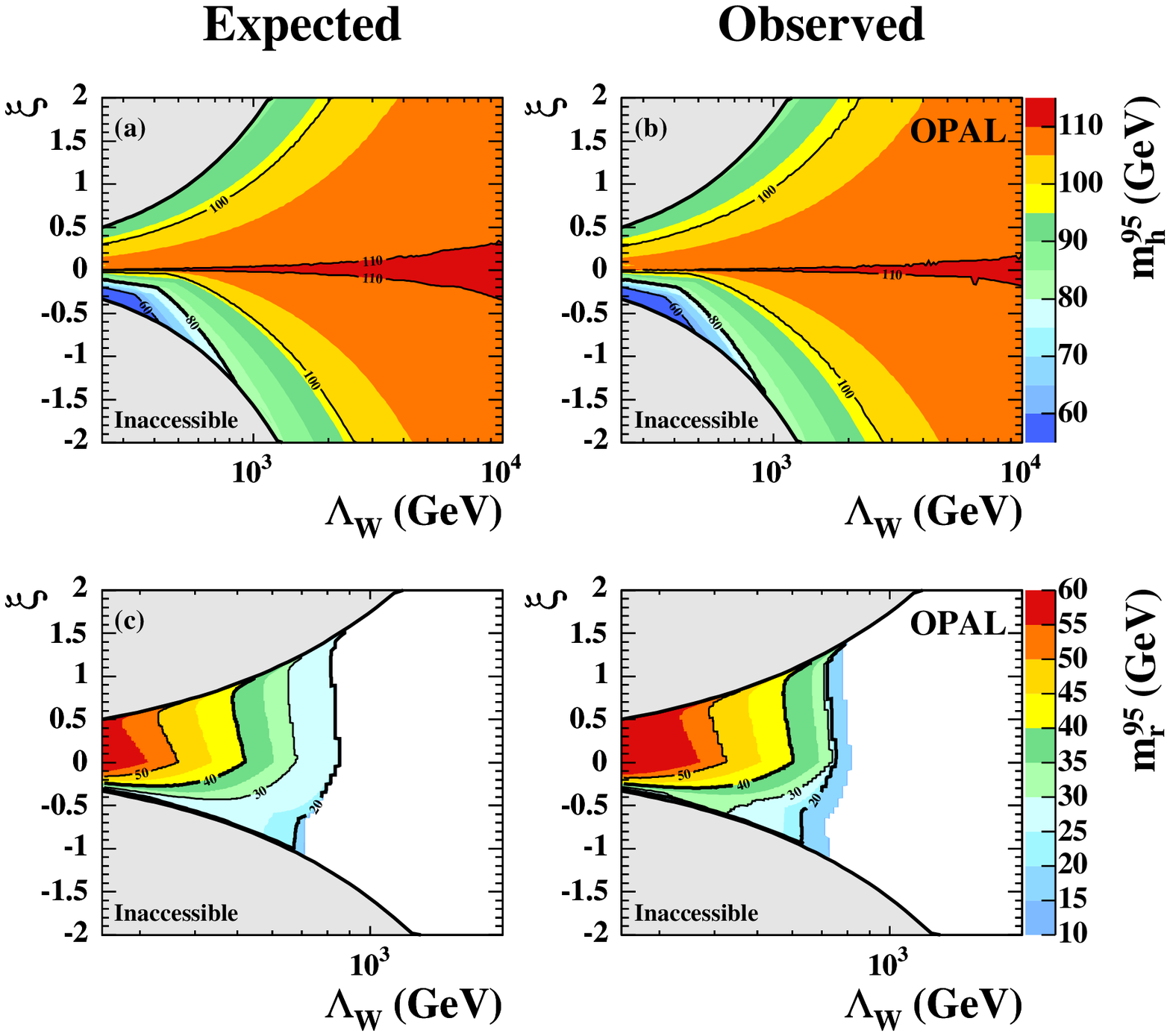} 
    \begin{picture}(0,0)
      \put(-160,0){\mbox{(b)}}
    \end{picture}
    \caption[*]{
      (a) Regions in the plane $(f, m_{\tilde{\pi}})$, excluded by the
      searches for $\ee\to\tilde{\pi}\tilde{\pi}+\gamma/\Z$;
      (b) radion-like state mass limits as a function of $\xi$ and
      $\Lambda_W$. 
      }
      \label{fig:brarad}
  \end{center}
\end{figure}

In addition to gravitons, the effective ADD four-dimensional theory of 
gravity predicts also the existence of {\em branons}
$\tilde{\pi}$, massive scalars related to the deformations of the
brane within the bulk that could be pair produced
($\ee\to\tilde{\pi}\tilde{\pi}+\gamma/\Z$), with cross section
depending on the number of branons $n_b$, their mass and the brane
tension $f$. A specific L3 search~\cite{L3bra} for such processes
allows branon mass limits to be set as a function of $f$, as shown
in Fig.~\ref{fig:brarad}(a).

In the Randall-Sundrum scenario two branes are assumed to exist: one
where SM is confined, one, the Planck brane, where the gravity is
confined. There is only one extra dimension and the gravity is
``weak'' because of its exponential suppression with the distance
between the SM and Planck branes. Fluctuations of this distance give
rise to a massive scalar, the {\em radion}, that can mix with the SM
Higgs bosons, having the same quantum 
numbers.  The resulting Higgs-like and a radion-like eigenstates can
be produced at LEP through the strahlung process and thus searched for
by using the Higgs boson selections. A result of these searches from
OPAL~\cite{OPALrad} is shown in Fig.~\ref{fig:brarad}(b), where the
limits on the radion-like state mass $m_r$ are given as a function
of the mixing parameter $\xi$ and the mass scale on the SM brane
$\Lambda_W$.

\section{Conclusion}
LEP has extensively searched for Higgs bosons of minimal and
non-minimal models, for sparticles within the most promising
supersymmetric scenarios and for many other possible new phenomena
beyond the Standard Model. More than the negative outcome, the wide
and detailed lessons learned in this challenge are the important part
of the LEP legacy that will result crucial for future experiments.

\section*{Acknowledgements} I would like to thank the organizers of
HS04 for their kind hospitality and for the opportunity to
give this nice review talk. 

\end{document}